\documentclass[twocolumn]{aastex631}

\defcitealias{2019MNRAS.490.1570B}{BR19}

\begin{document}

\title{First detection of silicon-bearing molecules in $\eta$ Car}

\author[0000-0002-7703-0692]{C. Bordiu}
\affiliation{INAF-Osservatorio Astrofisico di Catania, Via Santa Sofia 78, 95123 Catania}

\author[0000-0002-8443-6631]{J. R. Rizzo}
\affiliation{Centro de Astrobiología (INTA-CSIC), Ctra. M-108, km. 4, E-28850 Torrejón de Ardoz, Madrid, Spain}
\affiliation{ISDEFE, Beatriz de Bobadilla 3, E-28040 Madrid, Spain}

\author[0000-0002-3429-2481]{F. Bufano}
\affiliation{INAF-Osservatorio Astrofisico di Catania, Via Santa Sofia 78, 95123 Catania}

\author[0000-0002-5417-1943]{G. Quintana-Lacaci}
\affiliation{Group of Molecular Astrophysics. IFF. CSIC. C/ Serrano 123, E-28006, Madrid, Spain}

\author[0000-0002-7288-4613]{C. Buemi}
\affiliation{INAF-Osservatorio Astrofisico di Catania, Via Santa Sofia 78, 95123 Catania}

\author[0000-0003-4864-2806]{P. Leto}
\affiliation{INAF-Osservatorio Astrofisico di Catania, Via Santa Sofia 78, 95123 Catania}

\author[0000-0003-1856-6806]{F. Cavallaro}
\affiliation{INAF-Osservatorio Astrofisico di Catania, Via Santa Sofia 78, 95123 Catania}

\author{L. Cerrigone}
\affiliation{Joint ALMA Observatory, Alonso de C\'ordova 3107, Vitacura, 8320000, Santiago, Chile}

\author[0000-0002-3137-473X]{A. Ingallinera}
\affiliation{INAF-Osservatorio Astrofisico di Catania, Via Santa Sofia 78, 95123 Catania}

\author{S. Loru}
\affiliation{INAF-Osservatorio Astrofisico di Catania, Via Santa Sofia 78, 95123 Catania}

\author[0000-0001-6368-8330]{S. Riggi}
\affiliation{INAF-Osservatorio Astrofisico di Catania, Via Santa Sofia 78, 95123 Catania}

\author{C. Trigilio}
\affiliation{INAF-Osservatorio Astrofisico di Catania, Via Santa Sofia 78, 95123 Catania}

\author[0000-0002-6972-8388]{G. Umana}
\affiliation{INAF-Osservatorio Astrofisico di Catania, Via Santa Sofia 78, 95123 Catania}

\author[0000-0002-5574-2787]{E. Sciacca}
\affiliation{INAF-Osservatorio Astrofisico di Catania, Via Santa Sofia 78, 95123 Catania}

\begin{abstract}

We present ALMA band 6 observations of the luminous blue variable $\eta$ Car, obtained within the ALMAGAL program. We report SiO $J=5\rightarrow4$, SiS $J=12\rightarrow11$ and SiN $N=5\rightarrow4$ emission in the equatorial region of the Homunculus nebula, constituting the first detection of silicon- and sulphur-bearing molecules in the outskirts of a highly evolved, early-type massive star. SiO, SiS and SiN trace a clumpy equatorial ring that surrounds the central binary at a projected distance of $\sim$2 arcsec, delineating the inner rims of the butterfly-shaped dusty region. The formation of silicon-bearing compounds is presumably related to the continuous recycling of dust due to the variable wind regime of $\eta$ Car, that destroys grains and releases silicon back to gas phase. We discuss possible formation routes for the observed species, contextualizing them within the current molecular inventory of $\eta$ Car. We find that the SiO and SiS fractional abundances in localised clumps of the ring, $6.7\times10^{-9}$ and $1.2\times10^{-8}$ respectively, are exceptionally lower than those measured in C- and O-rich AGB stars and cool supergiants; while the higher SiN abundance, $3.6\times10^{-8}$, evidences the nitrogen-rich chemistry of the ejecta. These abundances must be regarded as strict upper limits, since the distribution of H$_2$ in the Homunculus is unknown. In any case, these findings shed new light onto the peculiar molecular ecosystem of $\eta$ Car, and establish its surroundings as a new laboratory to investigate the lifecycle of silicate dust in extreme astrophysical conditions.

\end{abstract}


\section{Introduction} \label{sec:intro}

$\eta$ Car is a massive binary system composed of a $\sim100$ M$_{\sun}$ luminous blue variable (LBV), and a hotter, $\sim$30 M$_{\sun}$ secondary with spectroscopic signatures typical of late-O and WR stars \citep{2005ApJ...633L..37I}.  The system is very eccentric (e = 0.9), with a well-determined orbital period of 5.54 years and a minimum separation at periastron of 1.5 A.U \citep{1996ApJ...460L..49D, 2008MNRAS.384.1649D}. The primary, $\eta$ Car A, is the archetypal member of the LBV family, and has captivated astronomers ever since it underwent a violent mass eruption in the 1840s \citep{2004JAD....10....6F}. This outburst, dubbed the "Great Eruption", released nearly $10^{50}$ erg \citep{2003AJ....125.1458S, 2013MNRAS.429.2366S} --rivalling the energetic output of a SN--, and ripped off between 20 and 40 M$_\sun$ of stellar material \citep{2017ApJ...842...79M}. The heavily processed ejecta, ashes of the CNO cycle, formed the Homunculus nebula \citep{1950ApJ...111..408G}, a rapidly expanding ($v_{\mathrm{exp}}\sim 650$ km s$^{-1}$, \citealt{1992A&A...262..153H,2018MNRAS.480.1466S}) hourglass-shaped circumstellar structure with a current-day projected size of $\sim0.18\times0.11$ pc. The Great Eruption was followed, almost 50 yr later, by a second, less energetic outburst \citep{2011MNRAS.415.2009S} that gave birth to an inner structure commonly referred to as the Little Homunculus \citep{2003AJ....125.3222I}, mimicking a gigantic matryoshka doll. The exceptionally complex circumstellar environment produced by these successive events makes of $\eta$ Car a unique testing ground to investigate (1) the nature of eruptive mass loss processes in massive stars near the Eddington limit \citep{2004ApJ...605..405S, 2019MNRAS.489..268S}; and (2) the intricate wind interactions in highly massive binary systems \citep{2008IAUS..250..133O, 2009MNRAS.396.1308G, 2016MNRAS.462.3196G}.

Located at 2350 pc from Earth \citep{2006ApJ...644.1151S} toward the near side of the Carina arm, $\eta$ Car and the Homunculus have been the target of multiple observing campaigns that, spreading across the entire electromagnetic spectrum, allowed for studying the central binary and its surroundings in great detail (see \citealt{davidson2012eta} for a review). Spectroscopic observations of the circumstellar material have provided an accurate portrait of the atomic gas and dust in and around the Homunculus, unveiling a chemical panorama quite in line with other LBV nebulae, i.e., nitrogen-rich ejecta with a significant oxygen depletion \citep{1982ApJ...254L..47D, 1986ApJ...305..867D, 1997ASPC..120..255D, 2004ApJ...605..854S}. However, due to the rather extreme conditions in the outskirts of $\eta$ Car --hot temperatures, strong FUV fields--, only in the last two decades has the molecular content of the Homunculus begun to draw attention. The first detection of neutral gas was achieved by \cite{2001ApJ...551L.101S}, who reported H$_2$ $v$ = 1 emission at 2.12 $\mu$m, tracing the surface of the Homunculus lobes. A few years later, high velocity absorption lines of CH and OH were identified through UV spectroscopy \citep{V2005ApJ...629.1034}, and shortly after ammonia was reported toward the core region of the Homunculus \citep{2006ApJ...645L..41S}. The first molecular survey was carried out by \cite{2012ApJ...749L...4L} with APEX, reporting six new species (CO, CN, HCO$^+$, HCN, HNC and N$_2$H$^+$) and two isotopologues ($^{13}$CO and H$^{13}$CN), thereby confirming $\eta$ Car as the LBV with the richest molecular inventory. 

Initial attempts to map the spatial distribution of this molecular material at high angular resolution with ATCA (HCN, \citealt{2016ApJ...833...48L}) and ALMA (CO, \citealt{2018MNRAS.474.4988S}) narrowed down the emission to the central few arcsec of the Homunculus. The gas, with a velocity dispersion of $\sim$300 km s$^{-1}$, has a kinematic age in good agreement with the Great Eruption, and follows the distribution of the equatorial structures of warm dust detected in mid-infrared observations \citep{1999Natur.402..502M}, tracing a sort of equatorial ring or torus of radius $\sim$4400 AU. This ring, disrupted towards the NW --its near side, coincident with the projected direction of the orbit's major axis--, delineates the bright rims of the so-called "butterfly" nebula \citep{2005A&A...435.1043C}, a region of efficient dust processing where emission from several other species has been resolved, such as HCN or HCO$^+$ (\citealt{2019MNRAS.490.1570B}, hereafter \citetalias{2019MNRAS.490.1570B}).  Interestingly, a thorough re-calibration of CO ALMA data by \cite{2022arXiv220513405Z}, significantly improving sensitivity, has revealed higher velocity components that may be tracing the walls of the polar lobes of the Homunculus by means of limb brightening.

The most recent observations toward $\eta$ Car with ALMA and Herschel continue enlarging its molecular inventory, reporting new detections of H$_2$O, CH$_3$OH \citep{2020ApJ...892L..23M} and multiple O-, C- and N-bearing species and reactive ions \citep{2020MNRAS.499.5269G}, many of them never detected before in a star of this kind. These findings unravel an unexpected molecular scenario for an early-type high-mass star in its final evolutionary stages. In this work, we analyse ALMA band 6 observations from the ALMAGAL program, with a resolution better than $\sim$0.4 arcsec, reporting the first detection of three Si-bearing molecules, namely SiO, SiS, and SiN, therefore adding yet another piece to the intriguing chemical puzzle of $\eta$ Car. In Sect. \ref{sec:observations} we describe the observations and data processing methods; in Sect. \ref{sec:results} we outline the findings of this work, putting them in a broader context and discussing their implications; and in Sect. \ref{sec:conclusions} we summarize the results and lay out the next steps in this research.

\section{Observations and data reduction} \label{sec:observations}

The ALMAGAL project (P.I: S. Molinari) is an ALMA Cycle 7 large program that aims at observing the 1 mm continuum and lines toward more than 1000 dense star forming clumps with $M > $500 M$_{\sun}$ and $d < 7.5$ kpc. In the context of this program, the field ID 653755, centred on the coordinates ($\alpha$, $\delta$) $=$ (10h45\arcmin03\farcs27, $-59$\degr41\arcmin03\farcs74), was observed. The field serendipitously covers the position of $\eta$ Car, located $\sim$2 arcsec East from the phase centre. Observations took place on April 4th, 2021, $\sim$one year after the last $\eta$ Car's periastron passage (on 2020.2), using 44 antennas with a maximum baseline of 1.4 km, resulting in a synthesized beam of $0.38\times0.31$ arcsec (P.A.: 1.38\degr). The field was observed for 45 min, under good weather conditions (precipitable water vapour $\sim$ 2.95 mm).

The spectral setup consisted of four spectral windows, two wide windows covering the frequency ranges 217.00..218.87 and 219.08..220.95 GHz (spectral resolution $\sim$ 976.56 KHz), and two narrower, overlapping windows covering the ranges 218.08..218.55 and 220.38.220.85 (spectral resolution $\sim$ 244.14 KHz). Raw visibilities were calibrated and imaged employing \texttt{CASA} (v6.1.1.15) with the ALMA pipeline version 2020.1.0.40. The final products were a 2D continuum map at a reference frequency of 220 GHz, and four spectral cubes (one for each spectral window). We manually reprocessed the products using the \texttt{SpectralCube}\footnote{\url{https://github.com/radio-astro-tools/spectral-cube}} Python library and the \texttt{GILDAS/CLASS}\footnote{\url{https://www.iram.fr/IRAMFR/GILDAS/}} software package. Despite the initial continuum subtraction performed on the visibilities before imaging, spectra toward the continuum-emitting region were found to have a highly structured baseline; this issue could be related to the amplification of small calibration systematics by the strong continuum of $\eta$ Car \citep{2022MNRAS.517...47A}. To deal with this problem, we fitted and subtracted a 3rd order polynomial to the  spectra in all positions. This method was able to remove a significant part of the spurious structure towards the central region, while keeping the baseline flat elsewhere. Finally, the data cubes were recentred on the J2000 coordinates of $\eta$ Car, ($\alpha$, $\delta$)=(10h45\arcmin03\farcs53, --59\degr41\arcmin04\farcs05).

In this work, we focus exclusively on the analysis of the cubes with the largest spectral coverage. Throughout the paper, we adopt the following conventions: (1) intensities are expressed in the original flux density scale of Jy beam$^{-1}$, unless explicitly noted otherwise; (2) velocities refer to the local standard of rest frame (LSR); and (3) positions are given as offsets from the coordinates of $\eta$ Car.

\section{Results and discussion} \label{sec:results}

\subsection{Detections and identification}

After a careful inspection of the cubes, we identified emission from five molecules, namely SiO, SiS, SiN, $^{13}$CO and $^{13}$CN. To our knowledge, this constitutes the first detection of Si- and S- bearing compounds in the outskirts of $\eta$ Car, and, in a wider context, in a highly evolved early-type massive star. Quantum numbers, lower level energies, and rest frequencies of the transitions detected are summarized in Table \ref{tab:mol_inventory}.

\begin{deluxetable}{llcc}[h!]
\tablecaption{List of molecules detected. \label{tab:mol_inventory}}
\tablecolumns{3}
\tablenum{1}
\tablewidth{0pt}
\tabletypesize{\scriptsize}
\tablehead{
\colhead{Species} &
\colhead{Transition} &
\colhead{$E_\mathrm{L}$ (cm$^{-1}$)} &
\colhead{$\nu_0$ (GHz)}
}
\startdata
SiO & $J=5\rightarrow4$ & 14.48 &  217.104 \\
SiS & $J=12\rightarrow11$ & 39.97 & 217.817  \\
SiN & $N=5\rightarrow4$, $J=9/2\rightarrow7/2$ & 14.52 & 218.006  \\
{ } & $N=5\rightarrow4$, $J=11/2\rightarrow9/2$ & 14.59 & 218.512 \\
$^{13}$CN & $N= 2\rightarrow1$, $J=3/2\rightarrow1/2$ & 3.62 & 217.303 \\
{ } & $N= 2\rightarrow1$, $J=5/2\rightarrow3/2$ & 3.65 & 217.467 \\
$^{13}$CO & $J=2\rightarrow1$ & 3.68 & 220.398 \\
\enddata
\end{deluxetable}

Silicon is an abundant element (solar abundance $n_\mathrm{Si}$/$n_\mathrm{H2}$ = $6.47\times10^{-5}$, \citealt{2021A&A...653A.141A}), most of which is depleted on to dust, being mainly stored in grain cores and, to a lesser extent, in the mantles \citep{2008A&A...482..809G}. Shocks can destroy dust grains either via ion sputtering or, especially when propagating in a dense medium ($n_\mathrm{H_2}>$10$^4$ cm$^{-3}$), via grain-grain collisions, effectively releasing Si back to gas phase \citep{1997A&A...321..293S,1997A&A...322..296C,2009A&A...497..145G,2011A&A...527A.123G}. For this reason, Si-bearing species like SiO are typically regarded as reliable tracers of shocked regions \citep{1992A&A...254..315M}. SiO and SiS have also been proposed as gas-phase precursors of silicate dust. In circumstellar envelopes of AGB stars, their fractional abundances have been observed to drop at a few stellar radii, possibly due to accretion onto dust grains \citep{1992A&A...262..491L,2006A&A...454..247S, 2007A&A...473..871S, 2016A&A...590A.127W}.

The study of Si-bearing molecules is therefore crucial to understand the lifecycle of dust. Both SiO and SiS have been widely searched for in many astrophysical environments: SiO has been detected in molecular clouds \citep{1971ApJ...167L..97W}, envelopes of evolved stars of all chemical types (e.g. \citealt{1979ApJ...229..257M,1994A&A...285..247B,2006A&A...454..247S}); star-forming regions \citep{1998ApJ...504L.109L,2004ApJ...603L..49J}; and even in the disk of the B[e] supergiant CPD-52 9243 \citep{2015ApJ...800L..20K}. Likewise, SiS has also been found in circumstellar envelopes of all chemical types \citep{1981A&A...101..238G, 2007A&A...473..871S, 2019MNRAS.484..494D, 2019A&A...628A..62M, 2020A&A...641A..57M}, star forming regions with outflows \citep{1981ApJ...247..112D,1988ApJ...324..544Z} and exceptionally, in L1157, a shocked region associated with a sun-like protostar \citep{2017MNRAS.470L..16P}. However, no detections have been achieved to date in the outskirts of a highly evolved, early-type massive star like $\eta$ Car, where the physical conditions are in principle much less hospitable.

Contrary to SiO and SiS, SiN is a rather elusive molecule, that to date has only been found in three astrophysical environments: the C-rich envelope of IRC$+$10 216 \citep{1992ApJ...388L..35T}, the hot molecular core SgrB2(M) \citep{2003A&A...412L..15S}, and the S-type AGB star W Aql \citep{2020A&A...642A..20D}.

Our line identification relies upon strong kinematic arguments, as only the SiO, SiS and SiN transitions listed in Table \ref{tab:mol_inventory} match the observed velocities of CO, HCN and HCO$^{+}$ \citepalias{2019MNRAS.490.1570B}. Still, we assessed other alternative identifications to account for the complex kinematics of the circumstellar material. Several transitions of CH$_3$OH and SO$_2$ lie within 100 MHz of the rest frequencies of SiO and SiS (217.104 and 217.817 GHz, respectively). However, we find their corresponding velocities to be largely incompatible with those of CO, HCN and HCO$^+$, in some cases off by more than $\sim$100 km s$^{-1}$. Moreover, the CH$_3$OH and SO$_2$ transitions have lower level energies ($E_\mathrm{L}$) in the range $\sim$500-2000 cm$^{-1}$, much higher than SiO and SiS, and also than the methanol transition detected by \cite{2020ApJ...892L..23M}. The excitation conditions required by such transitions are only achievable very close to the binary, where vibrationally excited HCN has been reported \citepalias{2019MNRAS.490.1570B}. 

Similarly, we searched for other candidate lines to explain the features at 218.006 and 218.512 GHz that we attribute to SiN. Both SiN components lie close in frequency to the 10$_{0,10}\rightarrow9_{0,9}$ and 9$_{2,7}\rightarrow8_{2,6}$ transitions of $^{29}$SiC$_2$ at 218.011 and 218.506 GHz. However, the non-detection of the 10$_{0,10}\rightarrow9_{0,9}$ transition of the presumably more abundant isotopologue SiC$_2$ at 220.773 GHz makes this identification highly unlikely, leaving SiN as the only plausible option.

\subsection{Spatial distribution and kinematics}\label{subsec:spatialkinematics}

\begin{figure*}[t!]
\figurenum{1}
\plotone{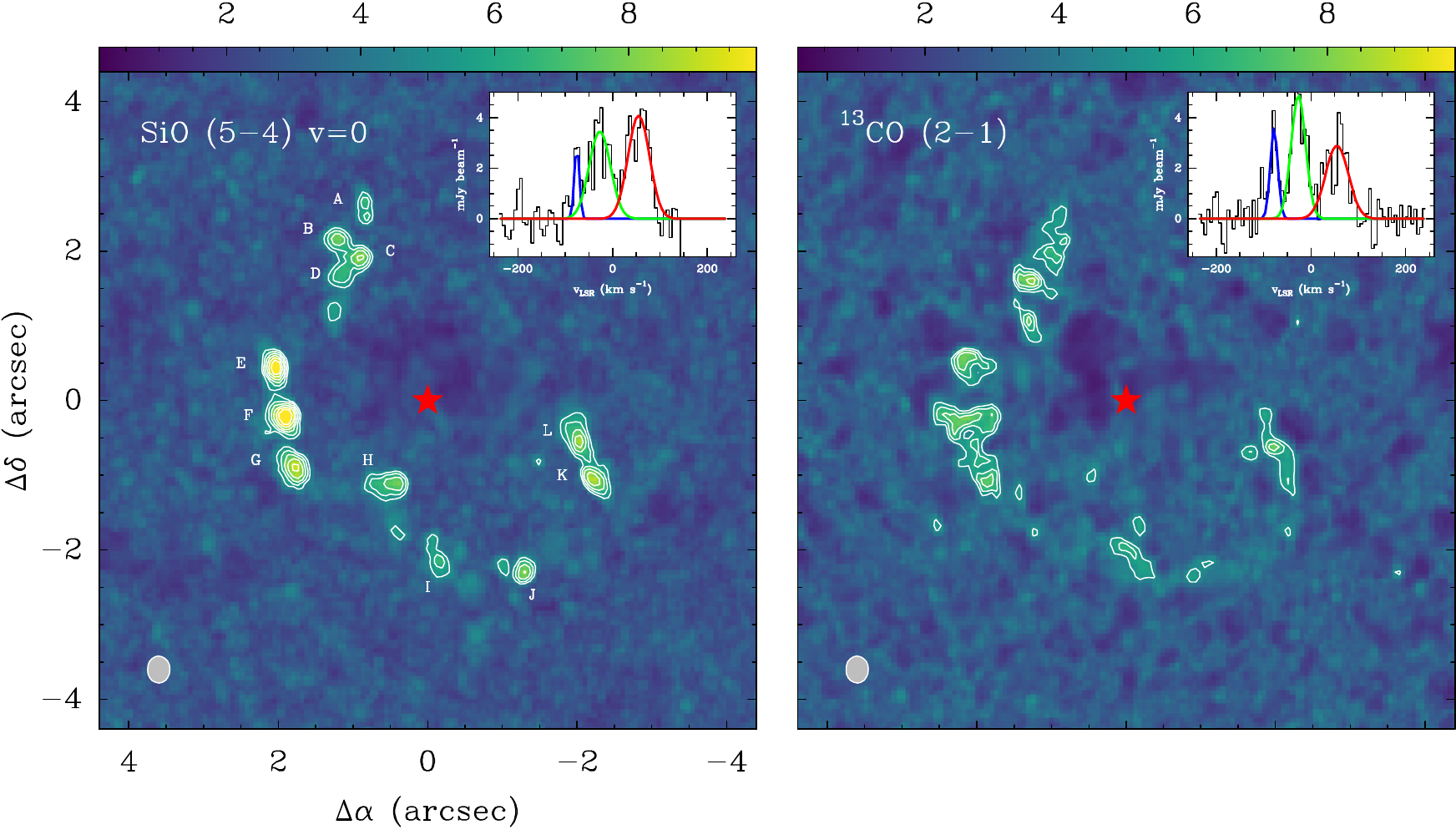}
\caption{S/N maps of SiO $J=5\rightarrow4$ ($v=0$) and $^{13}$CO $J=2\rightarrow1$ towards $\eta$ Car. Contours from 5 to 10. The position of $\eta$ Car is indicated by the red marker. The approximate beam size is shown in the bottom-left corner. The insets show the average spectrum of each molecule, computed in a region of radius 2.5 arcsec. The dominant kinematic components are highlighted with Gaussian profiles. Since SiO appears next to the edge of the band, its spectrum is truncated above 140 km s$^{-1}$. \label{fig:maps}}
\end{figure*}

All the molecules trace a clumpy "C-shaped" structure, following the overall spatial distribution of the equatorial ring already seen in CO, HCN, H$^{13}$CN and HCO$^+$ (see fig. 1 in \citetalias{2019MNRAS.490.1570B}). No molecular emission is detected in the direction of the star. Fig. \ref{fig:maps} shows the peak intensity maps of SiO and $^{13}$CO, the brightest lines in each band, divided by the corresponding rms maps. The resulting S/N maps highlight the genuine line emission, while removing spurious features toward the continuum-emitting region --where the noise is higher-- that affected the original peak intensity maps. For convenience, we identified twelve significant emission clumps above 5$\sigma$ in the SiO map. The clumps were labelled from A to L counter-clockwise starting from the North, as indicated in Fig. \ref{fig:maps}, and will henceforth serve as a positional reference. 

The SiO clumps have an average projected distance to the star of $\sim$2 arcsec. Despite their similar distribution along the butterfly region, SiO and $^{13}$CO are not perfectly correlated: closer inspection reveals an offset between emission peaks, especially pronounced in the SE/E clumps (the far side of the ring), the SiO being closer to $\eta$ Car in projection by $\sim$0.1--0.2 arcsec ($\sim275-550$ au). This offset, smaller than the beam, is genuine and was validated by checking against the CO $J=3\rightarrow2$ data presented in \citetalias{2019MNRAS.490.1570B}. It suggests that SiO may preferentially trace the inner rims of the butterfly nebula, more exposed to the fast wind of $\eta$ Car, as proposed by \cite{2005A&A...435.1043C}.

The spatial distribution of the other Si-bearing compounds is even more patchy, with reliable detections only in a subset of the clumps. SiS is detected towards clumps C, D, E, F, G, H, J and L, whereas SiN is only found in positions D, F, G and H. In clump F, the two transitions at 218.006 and 218.512 GHz are detected at a $\sim$3$\sigma$ level, whereas in the other clumps only the 218.512 GHz component is unambiguously found. Since the two transitions are expected to have similar line strengths, a relative difference could be explained either by noise fluctuations, or very localised non-LTE effects (as observed in NH$_3$ around G79.29+0.46, \citealt{2014A&A...564A..21R}). 

In any case, the low S/N of most of these detections prevents us from producing meaningful maps of SiS and SiN, as the lines are only clearly resolved in averaged spectra. In this sense, the non-detection of SiS and SiN in clumps where SiO is present may either be a matter of sensitivity, or imply an actual chemical differentiation. Likewise, the overall clumpiness of Si-bearing species may be due to the limited sensitivity of the data, or indicate that only in certain places the right physical conditions for their formation are met in presence of sufficient Si. In view of the different relative intensities and radial offset between SiO and $^{13}$CO, we are inclined to favour the second scenario.

The insets in Fig. \ref{fig:maps} show the spectra of SiO and $^{13}$CO, averaged in a circular aperture of radius 2.5 arcsec, centred on the star, that encompasses the main emission features. The two molecules show a similar overall profile, with most of the emission originating from the velocity range ($-$100, $+100$) km s$^{-1}$. We performed a Gaussian fitting, identifying three major kinematic components that explain the overall motion of the gas: blueshifted material towards the W/NW, the near side of the ring; redshifted gas towards the S/SE, the far side; and material towards the NE and SW moving at velocities relatively close to the systemic velocity of $\eta$ Car ($-19.7$ km s$^{-1}$, \citealt{2004MNRAS.351L..15S}). However, we note that the relative intensities of the components differ between SiO and $^{13}$CO. In SiO, the redshifted component is the strongest, as clumps in the far side (e.g. E, F, G) are significantly brighter than the others. Such a difference may be a consequence of local density enhancements, or modulations intrinsic to the orbital phase.

\begin{figure*}[ht!]
\figurenum{2}
\plotone{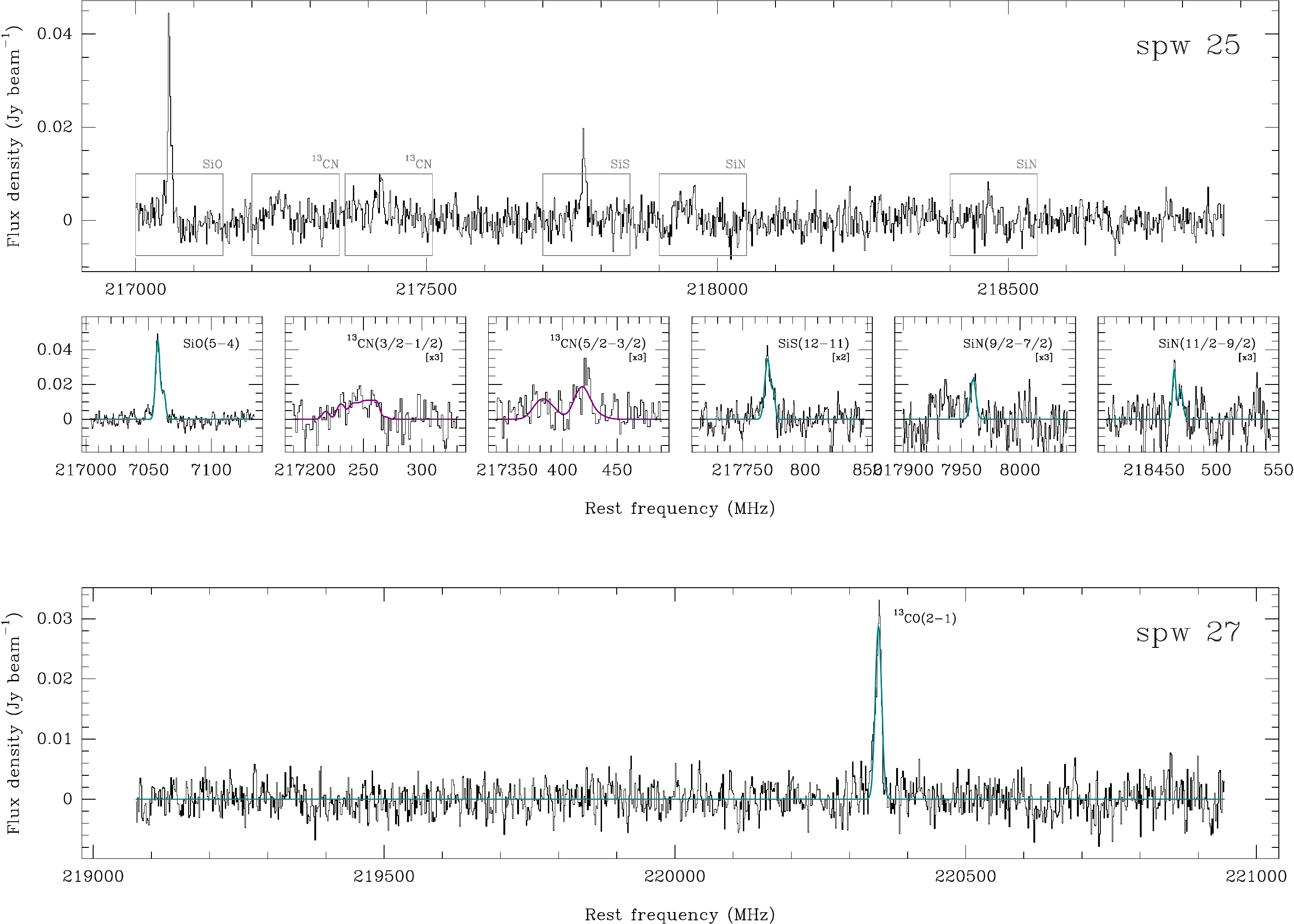}
\caption{Beam-averaged spectra of clump F. Top panel: full band from 217.0 to 218.8 GHz, rebinned to a resolution of 2 km s$^{-1}$. The insets show zoom-ins of the transitions detected (grey boxes on the full spectrum) at the original resolution, with a gaussian (blue) or hyperfine (purple) fitting superimposed. The $^{13}$CN spectra have been smoothed to better highlight the hyperfine structure. All the insets have the same flux density scale, with the different lines scaled appropriately for better visibility. Bottom panel: same as the top panel, from 219.1 to 221.0 GHz. \label{fig:spectrum}}
\end{figure*}

\begin{figure}[]
\figurenum{3}
\plotone{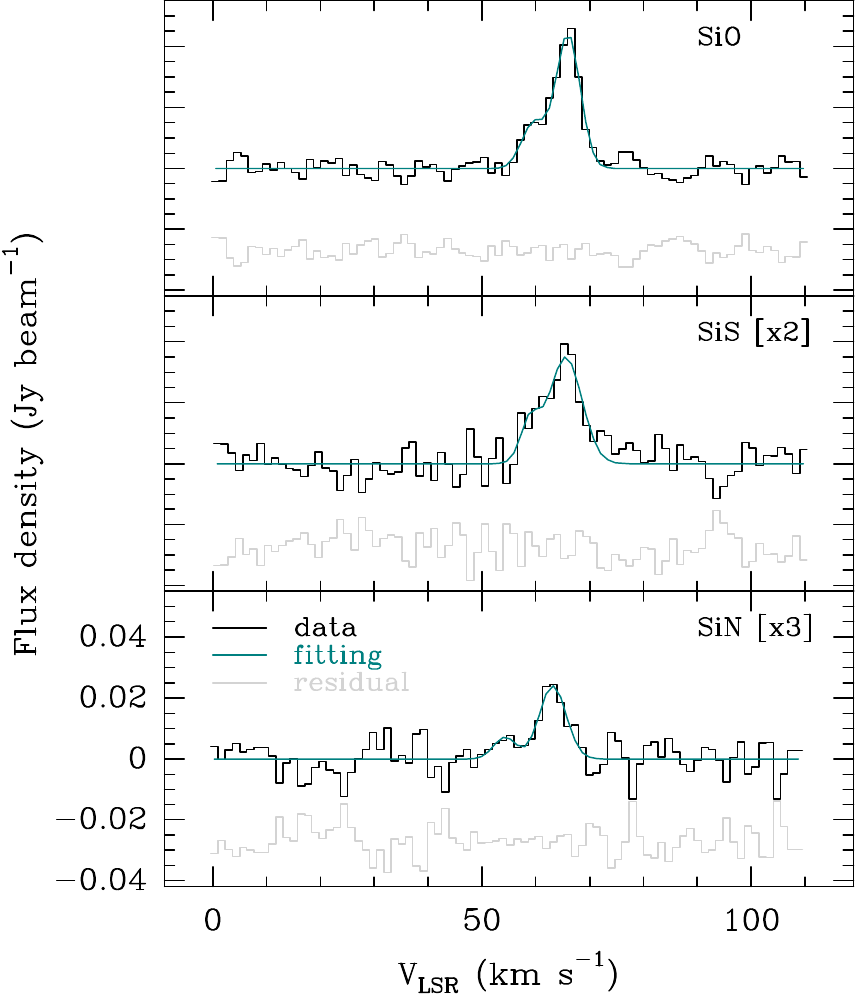}
\caption{Velocity-aligned spectra of the Si-bearing molecules detected in clump F. The SiN transitions at 218.006 and 218.512 GHz have been stacked to improve signal-to-noise ratio. For each spectrum, a gaussian fitting (blue) and the corresponding residual (gray) are shown. The relative scaling of the lines is the same as in Fig. \ref{fig:spectrum} \label{fig:spectrum-sibearing}}
\end{figure}

Individual clumps, on the other hand, exhibit a more complex kinematic structure, with multiple velocity components spreading over several tens of km s$^{-1}$. This dispersion is more significant in the clumps located south of the star, possibly tracing turbulent or stratified gas. For the subsequent analysis, beam-averaged spectra were extracted in all the twelve positions. Full spectra are presented in Appendix \ref{sec:full-spectra} (Figs. \ref{fig:fig-spw25-appendix} and \ref{fig:fig-spw27-appendix}). Fig. \ref{fig:spectrum} shows the spectra of clump F, the brightest one in SiO, as a representative example, while Fig. \ref{fig:spectrum-sibearing} presents a closer view of the spectra of the Si-bearing molecules in that position, aligned in velocity.

\subsection{Column densities \label{subsec:column-densities}}

We employed the software \texttt{MADCUBA}\footnote{MAdrid Data CUBe Analysis (MADCUBA), developed at the Centro de Astrobiolog\'ia. \url{https://cab.inta-csic.es/madcuba/}} \citep{2019A&A...631A.159M} to determine the column densities and molecular abundances of the observed species in all clumps. \texttt{MADCUBA} estimates the physical parameters of the emission by comparing the observed spectra with synthetic models computed under the only assumption of Local Thermodynamic Equilibrium (LTE). The \textsc{autofit} function of the \texttt{MADCUBA/SLIM} package (Spectral Line Identification and Modelling) performs a non-linear least-squares fit to the observed spectra, following the Levenberg-Marquardt algorithm \citep{Levenberg1944AMF,marquardt:1963}, and returns the set of parameters (column density $N$, excitation temperature $T_\mathrm{ex}$, line velocity and full width half maximum) that \text{best} reproduces the observed line profiles.

For the calculations, spectra were re-scaled to a main-beam temperature ($T_\mathrm{mb})$ scale, using:

\begin{equation}
    T_\mathrm{mb} = 1.222\times10^6\frac{S_\nu}{\nu^2\theta_\mathrm{maj}\theta_\mathrm{min}}
\end{equation}

\noindent with $S_\mathrm{\nu}$ the flux density in Jy beam$^{-1}$, $\nu$ the reference frequency in GHz, and $\theta_\mathrm{maj}$ and $\theta_\mathrm{min}$ the major and minor axes of the beam in arcsec.

With just a single rotational transition observed in all the lines, the excitation temperature cannot be determined. We thus have proceeded under the assumption of thermal coupling between dust and gas, setting the $T_\mathrm{ex}$ of the clumps to the black-body equilibrium temperature of dust grains, as given by \cite{2003AJ....125.1458S},

\begin{equation}
    T_\mathrm{BB} = 13100\,R_\mathrm{AU}^{-1/2}\, \mathrm{K}
\end{equation}

\noindent where $R_\mathrm{AU}$ is the physical distance from each clump to $\eta$ Car in AU, deprojected assuming an inclination of 41$^\circ$ \citep{2006ApJ...644.1151S}. This approach yields temperatures in the range 150--200 K, similar to those typically assumed in the literature. Finally, since all the clumps have a size comparable to the beam, we did not apply any filling factor correction.

Table \ref{tab:column-densities} presents the column densities of SiO, SiS, SiN, $^{13}$CO and $^{13}$CN in the clumps, with their associated uncertainties. When a molecule is not detected, the $3\sigma$ upper limit of the column density is provided. As discussed in Sect. \ref{subsec:spatialkinematics}, some clumps show a complex kinematic structure, which results in multiple resolved velocity components. For those, the sum of independent column densities along the line of sight is provided. Line fitting parameters (observed velocity $v$, full width half maximum $\Delta v$, and integrated intensity) for all the components with a signal-to-noise ratio above $3\sigma$ are listed in Appendix \ref{sec:line-fitting} (Tables \ref{tab:fitting-SiOSiSSiN} and \ref{tab:fitting-COCN}). 

\begin{deluxetable*}{lllccccc}[t]
\tablecaption{Column densities of SiO, SiS, SiN, $^{13}$CO and $^{13}$CN in the positions indicated in Fig. \ref{fig:maps}. \label{tab:column-densities}}
\tablecolumns{8}
\tablenum{2}
\tablewidth{0pt}
\tabletypesize{\scriptsize}
\tablehead{
\colhead{Pos} &
\colhead{($\Delta\alpha$, $\Delta\delta$)} &
\colhead{$T_\mathrm{ex}$} &
\colhead{$N(\mathrm{SiO})$} &
\colhead{$N(\mathrm{SiS})$} &
\colhead{$N(\mathrm{SiN})$} &
\colhead{$N(^{13}\mathrm{CO})$} &
\colhead{$N(^{13}\mathrm{CN})$} \\
\colhead{} & \colhead{} & \colhead{(K)} &
\colhead{($\times10^{14}$ cm$^{-2}$)} &
\colhead{($\times10^{14}$ cm$^{-2}$)} &
\colhead{($\times10^{14}$ cm$^{-2}$)} &
\colhead{($\times10^{17}$ cm$^{-2}$)} &
\colhead{($\times10^{15}$ cm$^{-2}$)}
}
\startdata
A & ($+0.9$, $+2.6$)   & 154 & 0.74 (0.07) & $<0.64$ & $<1.46$          & 0.69 (0.07) & $<0.14$ \\
B & ($+1.2$, $+2.2$)   & 166 & 1.53 (0.21) & $<0.51$ & $<1.51$          & 1.19 (0.61) & $<0.15$ \\
C & ($+0.9$, $+1.9$)   & 180 & 1.42 (0.11) & 2.25 (0.12) & $<1.65$      & 1.34 (0.13) & $<0.42$ \\
D & ($+1.2$, $+1.7$)   & 185 & 1.39 (0.08) & 2.19 (0.26) & 7.11 (6.89)  & 2.61 (0.09) & 0.88 (0.15) \\
E & ($+2.0$, $+0.4$)   & 179 & 1.79 (0.05) & 3.38 (0.41) & 	$<1.62$     & 1.76 (0.10) & 0.73 (0.15) \\
F & ($+2.0$, $-0.2$)   & 173 & 2.75 (0.24) & 4.83 (0.85) & 17.38 (7.85) & 2.44 (0.19) & 0.92 (0.15) \\
G & ($+1.8$, $-0.9$)   & 163 & 2.51 (0.50) & 3.38 (0.24) & 7.46 (0.85)  & 1.74 (0.46)  & $<0.42$      \\
H & ($+0.5$, $-1.1$)   & 211 & 3.51 (0.39) & 7.05 (1.30) & $<2.69$      & 2.89 (0.25) & 1.17 (0.34) \\
I & ($-0.3$, $-2.2$)   & 165 & 2.90 (0.33) & $<0.40$     & $<1.58$      & 3.45 (0.32) & $<0.25$      \\
J & ($-1.3$, $-2.3$)   & 162 & 1.33 (0.07) & 3.56 (0.36) & $<1.54$      & 3.19 (0.15) & $<0.17$       \\
K & ($-2.3$, $-1.1$)   & 168 & 2.04 (0.16) & $<0.53$     & 	$<2.02$     & 2.40 (0.15) & 1.34 (0.20) \\
L & ($-2.0$, $-0.5$)   & 180 & 1.66 (0.07) & 1.30 (0.20) & $<2.20$      & 2.25 (0.10) & 1.34 (0.24) \\
\enddata
\tablecomments{For clumps with multiple velocity components (see Tables \ref{tab:fitting-SiOSiSSiN} and \ref{tab:fitting-COCN}), we provide the sum of column densities of individual components. When a line is not detected, we provide the 3$\sigma$ upper limit of the column density.}
\end{deluxetable*}

We also checked for non-LTE effects using \texttt{RADEX} \citep{2007A&A...468..627V}. \texttt{RADEX} solves the statistical equilibrium equations using the escape probability formulation \citep{1960mes..book.....S}, without assuming LTE conditions. For the calculation, we set the H$_2$ volume density to 10$^{8}$ cm$^{-3}$, as suggested by \cite{2020MNRAS.499.5269G}, the kinetic temperature to the temperatures in Table \ref{tab:column-densities}, and the column density to the values provided by \texttt{MADCUBA} for each component. Then, we compared the integrated line intensities observed with those predicted by \texttt{RADEX}.  We found an average agreement better than 85\% for SiO, better than 95\% for SiS, and better than 97\% for $^{13}$CO. The slight discrepancy in SiO could perhaps be attributed to a possible weak maser contribution (see Sect. \ref{subsec:maser}). In any case, considering the assumptions made, we conclude that non-LTE effects, while present, might not be particularly significant for the measured transitions.

All the lines are found to be optically thin, with $\tau$ in the range (0.01--0.1). When detected, the column densities range roughly from 0.7 to $3.5\times10^{14}$ cm$^{-2}$ for SiO, from 1.3 to $7.1\times10^{14}$ cm$^{-2}$ for SiS, and from 0.7 to $1.7\times10^{15}$ cm$^{-2}$ for SiN. We observe column density variations among clumps up to a factor of $\sim$5 in the case of SiO and SiS. Considering only the positions where the two species are detected (C, D, E, F, G, H, J and L), the [SiO/SiS] ratio has an average value of 0.6, meaning a slight overabundance of SiS over SiO. The clumps where SiS is not detected, namely A, B, I and K, have much higher ratios when considering the $N(\mathrm{SiS})$ upper limits. Since the rms is similar across the clumps, the strong detections of SiO in absence of SiS in these positions may indicate a chemical differentiation. It is noteworthy that these clumps are among those located farthest from the star and hence with the lowest $T_\mathrm{ex}$. Contrarily, SiN is detected only towards four clumps (one of which is a tentative detection, see Table \ref{tab:fitting-SiOSiSSiN}), showing the highest column density of the Si-bearing species. These clumps are not among those with the highest SiO and SiS column densities, which could possibly imply a localised overabundance of SiN.

\subsection{Molecular abundances}\label{subsec:abundances}

\begin{figure}[ht!]
\figurenum{4}
\includegraphics[width=\columnwidth]{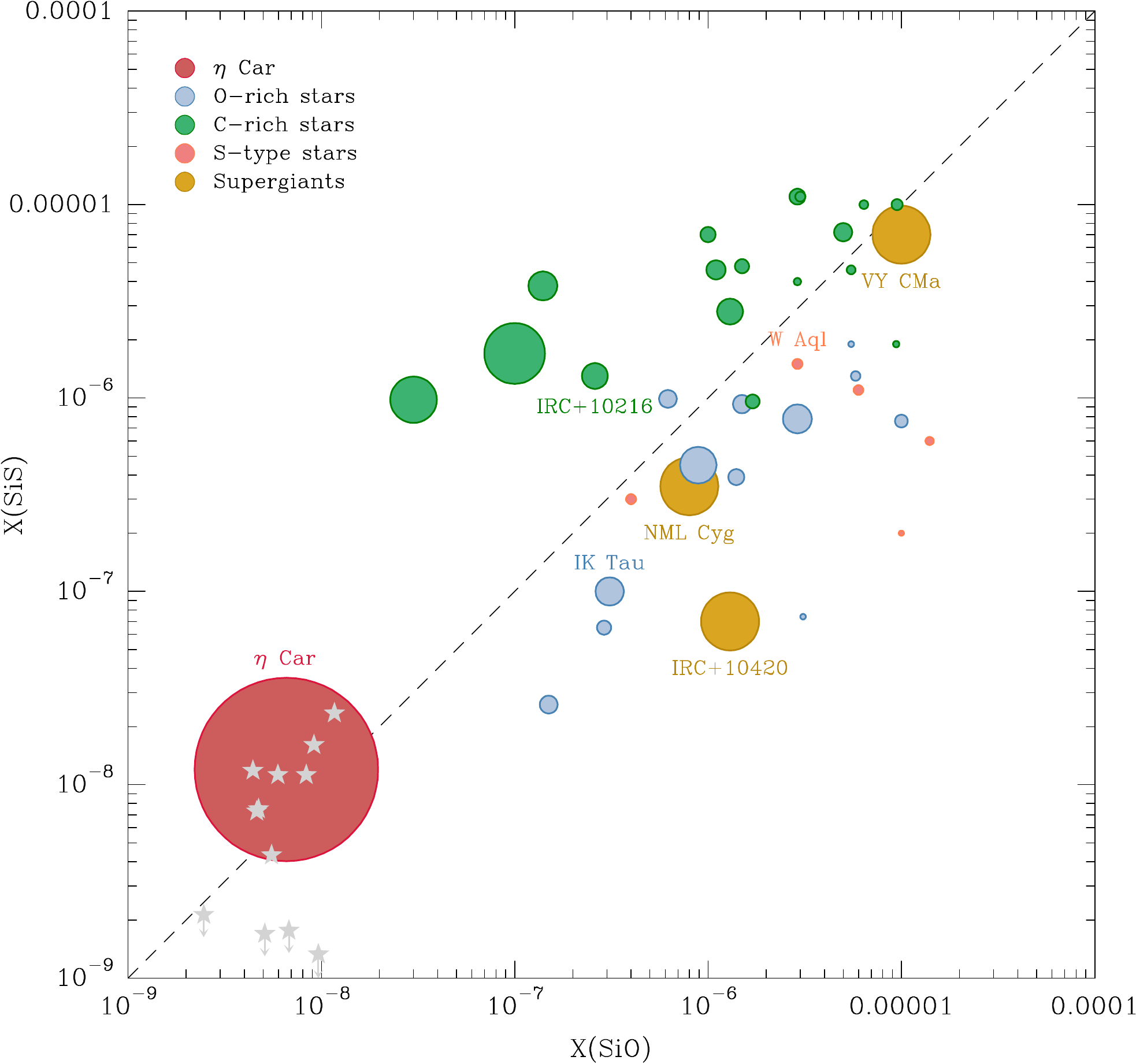}
\caption{Comparison of abundances of SiO and SiS measured in $\eta$ Car (this work) and other star types with confirmed detections of the two species: C-rich stars \citep{2019A&A...628A..62M}, O-rich stars \citep{2020A&A...641A..57M}, S-type stars \citep{2009A&A...499..515R,2018A&A...617A.132D,2020A&A...642A..20D}, the yellow hypergiant IRC+10420 \citep{2016A&A...592A..51Q}, and the red supergiants VY CMa \citep{2007Natur.447.1094Z} and NML Cyg \citep{2021ApJ...920L..38S}. Some relevant objects have been labelled for clarity. The size of the markers is proportional to the square root of $\dot M$. The gray star markers around $\eta$ Car indicate the abundances of individual clumps. The dashed line represents [SiO/SiS]$=1$.\label{fig:abundances}}
\end{figure}

Translating column densities into molecular abundances is not straightforward in $\eta$ Car, as the distribution of H$_2$ at angular scales comparable to the clumps is unknown. We are thus forced to adopt the global value of $N$(H$_{2}$) = $3.0 \times 10^{22}$ cm$^{-2}$ determined by \cite{2006ApJ...644.1151S}. Taking average column densities for each molecule, considering only clumps with detections above $3\sigma$, we derive [SiO/H$_\mathrm{2}$] = 6.7$\times$10$^{-9}$, [SiS/H$_\mathrm{2}$] = 1.2$\times$10$^{-8}$ and [SiN/H$_\mathrm{2}$] = 3.6$\times10^{-8}$. These values have to be regarded as peak abundances or upper limits, as we are assuming a homogeneous distribution of H$_\mathrm{2}$, which may not reflect reality, i.e., it is likely that H$_\mathrm{2}$ is somewhat enhanced in the clumps, resulting in lower abundances. Nevertheless, these values are a useful first estimate that allows us to put $\eta$ Car in context with other evolved stars with a rich silicon chemistry.

Fig. \ref{fig:abundances} compares the SiO and SiS abundances of $\eta$ Car with those measured in other sources, namely C-rich stars \citep{2019A&A...628A..62M}, O-rich stars \citep{2020A&A...641A..57M}, S-type stars \citep{2009A&A...499..515R,2018A&A...617A.132D,2020A&A...642A..20D}, the red supergiants VY CMa \citep{2007Natur.447.1094Z} and NML Cyg \citep{2021ApJ...920L..38S}, and the yellow hypergiant IRC+10420 \citep{2016A&A...592A..51Q}. Only stars with confirmed detections of the two species have been included. We find that the abundances of $\eta$ Car are exceptionally low, more than one order of magnitude lower than in other sources. This is in excellent agreement with the trend observed by \cite{2003A&A...411..123G}, who surveyed a large sample of AGB stars and found that the SiO fractional abundance and the mass-loss rate were inversely correlated regardless of the chemical type of the envelope, possibly as a consequence of enhanced depletion of SiO onto dust grains at higher densities. The case of SiS is particularly interesting. It is not detected in envelopes with $\dot M<10^{-6}$ M$_\odot$ yr$^{-1}$, possibly due to a lack of available Si and S. Above this threshold, \cite{2019A&A...628A..62M,2020A&A...641A..57M} found weak hints of an abundance decrease with increasing $\dot M$, similar to SiO, but only in C-rich envelopes. However, the abundances of $\eta$ Car, with its extreme mass-loss rate of 10$^{-3}$ M$_\odot$ yr$^{-1}$ \citep{2001ApJ...553..837H}, and other sources with mass-loss rates of a few $\sim10^{-4}$ M$_\odot$ yr$^{-1}$, such as NML Cyg and IRC+10420, seem to support this trend. The only exception is VY CMa, which presents large abundances of SiO and SiS despite having a high mass-loss rate ([SiO/H$_2$] = 10$^{-5}$, [SiS/H$_2$] = 7$\times10^{-6}$,  \citealt{2007Natur.447.1094Z}).

Besides, in $\eta$ Car, the clumps with detection of SiO and SiS have [SiO/SiS] $<$ 1 (except for clump L), meaning that their silicon chemistry is somewhat closer to that of C-rich stars, where SiS tends to be more abundant than SiO by a factor of a few. Such an abundance ratio is not immediately explainable, considering: 1) the composition of the ejecta around $\eta$ Car, which is rich in N but C- and O-poor; and 2) the lack of reported observations of sulphur-bearing species to explain the formation of SiS (see Sect. \ref{subsec:formation}). In this respect, \cite{2018ApJ...862...38Z} developed a simple shock model that predicts a slower post-shock formation of SiS than SiO, and with lower abundances. Still, we emphasize that $\eta$ Car is an exceptional environment with rather unusual physical conditions, that probably will require the development of specific models to properly describe its molecular chemistry. In this context, a way to explain the prevalence of SiS over SiO follows the arguments presented by \cite{2019A&A...628A..62M}, who regard SiS as a more efficient gas-phase reservoir of Si, to the detriment of SiO, which is more easily incorporated back to dust grains as density and mass-loss rate increase. Indeed,  condensation of SiO -and, to a lesser extent, of SiS-- onto dust could be a recurring process in the molecular ring. The material in the inner rims of the butterfly region would be constantly recycled, with shocks caused by the binary winds periodically destroying dust and leading to the formation of Si-bearing species. A fraction of these molecules would eventually deplete back to dust, in a cycle modulated by the orbital period of 5.54 yr.

Finally, it is worth noting that photodissociation could also play an important role to explain the observed abundances. Unfortunately, its impact is difficult to calibrate: first, because the UV flux to which the circumstellar material is exposed is, again, a function of the orbital phase: depending on the epoch, a large fraction of the ionizing flux from the hotter secondary is blocked by the optically thick primary wind \citep{2016MNRAS.462.3196G}; and second, because the photodissociation rate of SiS is not yet well determined, and usually assumed similar to that of SiO \citep{2019A&A...628A..62M}.

\subsection{Chemistry of Si-bearing molecules in the Homunculus}\label{subsec:formation}

The physics of dust formation in the massive winds and eruptions of hot evolved stars were studied by \cite{2011ApJ...743...73K}, who found that only extremely high mass-loss rates ($\dot M > 10^{-2.5}$ M$_\odot$ yr$^{-1}$) allowed for efficient condensation while providing sufficient self-shielding from UV radiation. In this context, the supply of gas-phase Si in $\eta$ Car could be explained by looking at the impact of the fast winds of the central stars on the ejecta from the Great Eruption. In timescales of months/years, several solar masses of processed material were expelled, a fraction of which expanded in the equatorial plane of the system. This matter, dense and hot, eventually cooled down enough as to allow for the condensation of dust grains.

The dust content of the Homunculus at different spatial scales has been thoroughly studied by \cite{1999Natur.402..502M} and \cite{2005A&A...435.1043C}, who concluded that only a mix of silicates and corundum can account for the spectral features observed in the infrared. This dust has been continuously exposed to the strong, optically thick primary wind (v$_{\inf} \sim$ 420 km s$^{-1}$), plus the intermittent impact of the much faster secondary wind (v$_{\inf} \sim$3000 km s$^{-1}$) modulated by the orbital cycle. The competing effect of these winds could possibly explain the processing of dust in the inner borders of the butterfly region, releasing Si back to gas phase (and explaining the radial offset observed between SiO and CO/$^{13}$CO emission reported in Sect. \ref{subsec:spatialkinematics}). We could also speculate about the differences in brightness of the SiO clumps resorting to the most accepted orbital geometry (e.g. \citealt{2016ApJ...819..131T}), which places the periastron on the far side of the orbit (i.e. "behind" $\eta$ Car A, in the S/SE direction where SiO emission is more intense). This is also the dominant direction where receding, highly compressed fossil wind structures pile up \citep{2016MNRAS.462.3196G} as a result of the fast dive of the secondary into the primary wind. The accumulated effect of such abrupt changes in the wind regime over several orbital cycles may favour the dust processing on the far side, hence explaining the observed brightness differences.

Chemical models addressing the formation and destruction of Si-bearing species predict two dominant formation routes for SiO, namely: (1) oxidation of gas-phase Si, primarily through Si + CO $\rightarrow$ SiO + C and Si + OH $\rightarrow$ SiO + H \citep{2006ApJ...650..374A}, and (2) reaction with sulphur-bearing compounds, via Si + SO $\rightarrow$  SiO + S, and Si + SO$_2$ $\rightarrow$ SiO + SO \citep{2018ApJ...862...38Z}. The detection of CO \citep{2012ApJ...749L...4L, 2018MNRAS.474.4988S} and OH in the Homunculus \citep{V2005ApJ...629.1034, 2020MNRAS.499.5269G} may suggest that the former reactions are the preferential formation channels of SiO, whereas no other sulphur-bearing species --besides SiS, in this work-- have been ever reported. In particular, SO$_2$, one of the main reservoirs of sulphur in AGB stars \citep{2016A&A...588A.119D}, has not yet been detected towards $\eta$ Car. The ALMAGAL setup covers eight transitions of SO$_2$ spanning a wide range of $E_\mathrm{L}$ values ($\sim$200--3000 cm$^{-1}$), allowing us to set a $3\sigma$ upper limit column density $N$(SO$_2$)$<$10$^{18}$ cm$^{-2}$ following the assumptions in Sect. \ref{subsec:column-densities}. This value, however, does not really set tight constraints, and still allows for a high fractional abundance. Finally, an alternate pathway for SiO formation involves the destruction of SiS through SiS + O $\rightarrow$ SiO + S. For this reaction to be dominant, sufficient free oxygen has to be available, which is unlikely in the environment of $\eta$ Car, given that most of it should be already locked in CO, and other less abundant species like OH. However, it could explain an overabundance of SiO in the clumps where no SiS is detected.

Regarding SiS, the kinetics of chemical reactions leading to its formation are less understood. The primary paths involve reactions with SO and SO$_2$, competing with SiO but having lower reaction rate coefficients \citep{2018ApJ...862...38Z}. As a consequence, the formation rate of SiO is always higher (by a factor of 5--7) than SiS. In a recent work, \cite{2021ApJ...920...37M} proposed an alternate gas-phase reaction, Si + SH $\rightarrow$ SiS + H, that was able to successfully explain the SiS overabundances observed in L1157 \citep{2017MNRAS.470L..16P}. 

Follow-up observations specifically targeting other sulphur carriers will provide better insight on the feasibility of the different formation routes of SiO and SiS. We note, though, that most of the available chemical models address the conditions and abundances found in circumstellar envelopes or star forming regions, and hence may not be directly applicable to $\eta$ Car, where conditions may differ by a large margin. The development of \textit{ad hoc} models will surely be needed to explain the lifecycle of Si-bearing species in such a harsh, unusual ambient.

The formation of SiN is probably related to that of SiO and SiS. However, the current understanding of SiN gas-phase chemistry is  limited. Provided that shocks supply enough Si, one of the main formation routes for SiN would be the neutral-neutral reaction \citep{1988A&A...199..127R} Si + NH $\rightarrow$ SiN + H. In C-rich environments, another possible source of SiN could be a dissociative electron recombination following the reaction Si$^+$ + NH$_3$ $\rightarrow$ SiNH$_2^+$ + H \citep{1992ApJ...388L..35T}. In $\eta$ Car, these two routes seem feasible, considering the recent detection of NH and NH$_3$ \citep{2020MNRAS.499.5269G}, a significant fraction of which may reside in the equatorial regions where SiN is found. Indeed, the high SiN abundance reflects the unusually nitrogen-rich chemistry of the Homunculus, consistent with heavily processed CNO ejecta.

\subsection{Is maser emission possible?\label{subsec:maser}}

\begin{figure}[ht!]
\figurenum{5}
\includegraphics[width=\columnwidth]{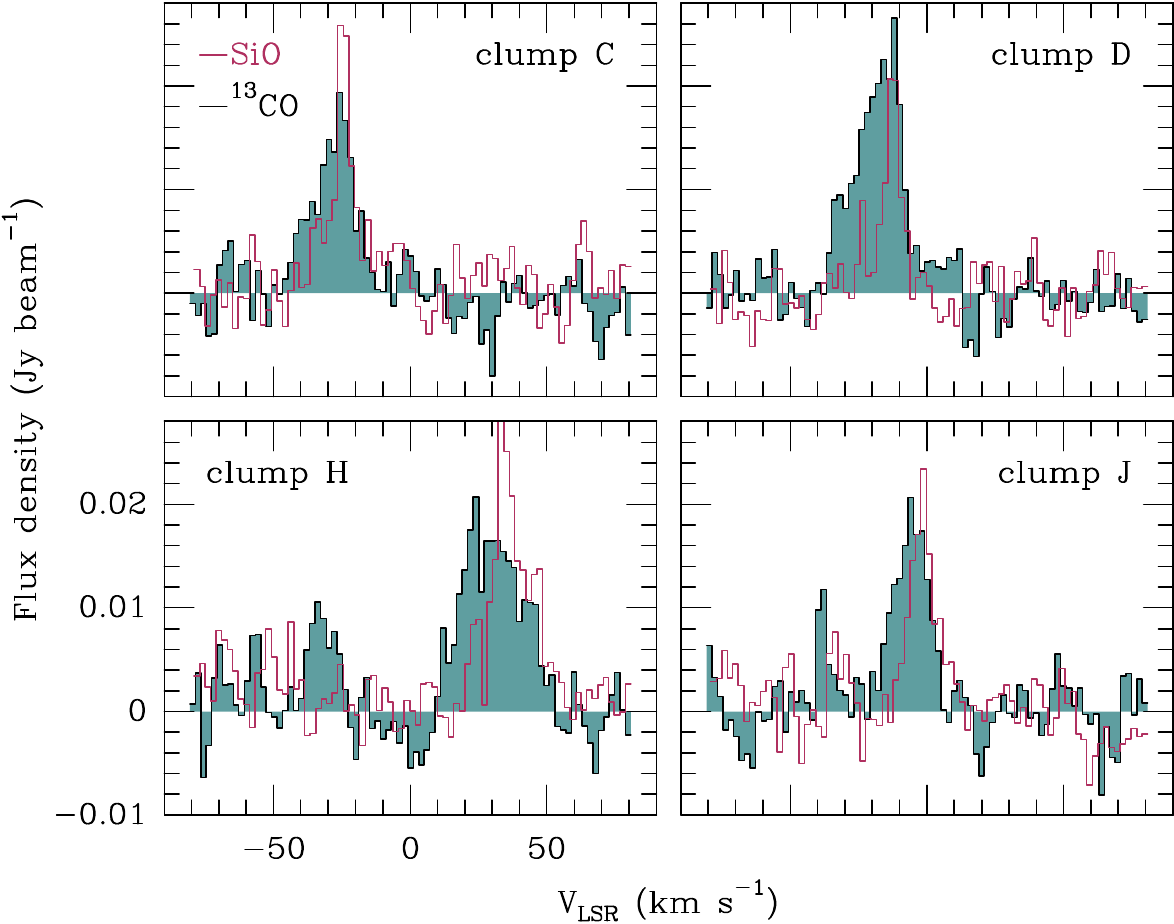}
\caption{Comparison of the SiO and $^{13}$CO line profiles in clumps C, D, H and J. Spectra have been smoothed to 2 km s$^{-1}$. All panels have the same intensity scale. \label{fig:profiles}}
\end{figure}

The column densities and abundances derived in Sect \ref{subsec:abundances} assume a thermal origin for the emission of all molecules. However, masers of SiO and its isotopologues have been frequently observed in envelopes of evolved stars \citep{1982ApJ...256L..55S, 1998A&A...329..219P, 2010ApJS..188..209K, 2021ApJS..253...44R}, even in their vibrational ground states ($v=0$). SiS masers have been detected as well, but only in IRC+10216 \citep{1983ApJ...267..184H,2006ApJ...646L.127F}. One may wonder whether the maser scenario may also hold in $\eta$ Car. In fact, SiO and SiS show some features that might be compatible with a maser contribution. First, SiO maser emission in evolved stars is generally clumpy, arising from discrete, very compact regions that usually trace a circumstellar ring due to tangential amplification \citep{1992A&A...254L..17C, 1994ApJ...430L..61D, 1997VA.....41..175D}; such a morphology resembles the spatial distribution of SiO around $\eta$ Car, although at very different angular scales. Second, maser emission tends to show narrower velocity profiles than thermal emission; in our case, we observe that SiO and SiS have significantly narrower lines than $^{13}$CO in many of the studied positions, as shown in Fig. \ref{fig:profiles}.

The pumping mechanism of SiO masers is not completely established, with radiative and collisional pumping as the most plausible explanations. For these mechanisms to be effective, H$_2$ densities in excess of $\sim$10$^9$ cm$^{-3}$ and kinetic temperatures above 1000 K are required \citep{1992ASSL..170.....E, 1994A&A...285..953B}. Such stringent physical conditions are typically found at a few stellar radii from the stellar photosphere in the case of AGB stars, hence posing an important drawback for $\eta$ Car, where material is located $\sim5000$ AU away from the binary. However, at this point there is no ground for rejecting the possibility that these conditions are locally met in unresolved, hot and dense pockets of gas. Any further assumption is mere speculation, as we cannot be certain about the nature of the SiO and SiS lines with just a single transition observed. Follow-up interferometric observations of higher $J$ transitions and other vibrational states, will elucidate whether the observed emission is entirely thermal, or has a weak maser contribution.

\section{Summary and Conclusions\label{sec:conclusions}}

We have analysed ALMA band 6 observations toward $\eta$ Car with a resolution better than 0.4 arcsec. We report the detection of SiO, SiS and SiN, the first Si- and S-bearing species found in the outskirts of a highly evolved, early-type massive star. The molecules trace a clumpy equatorial ring that surrounds the central binary at a projected separation $\sim$2 arcsec, presumably dating back to the Great Eruption in the 19th century.

Considering the abrupt changes of the wind regime due to the binary interaction, we propose the continuous processing of dust grains in the inner rim of the butterfly region as the most likely source of the gas-phase silicon needed for the formation of these species. The recent detection of other molecules typically associated with shock-heated gas, like NH$_3$, CH$_3$OH, H$_2$O or CH$^{+}$ \citep{2020ApJ...892L..23M,2020MNRAS.499.5269G}, even when their exact location in the Homunculus is yet to be pinpointed, may support this interpretation.

We have derived molecular abundances of [SiO/H$_2$] = $6.7\times10^{-9}$, [SiS/H$_2$] = 1.2$\times$10$^{-8}$, and [SiN/H$_2$] = 3.6$\times$10$^{-8}$ in the clumps. The SiO and SiS abundances are significantly lower than those measured in AGB stars and cool supergiants, (1) making of $\eta$ Car an outstanding object from the perspective of silicon chemistry, and (2) further confirming the trend of decreasing SiO and SiS abundances with increasing $\dot M$ inferred from observations of AGB stars. We suggest a more efficient re-adsorption of SiO into dust grains to explain the observed [SiO/SiS] ratio, whereas the prevalence of the rarer SiN over the other Si-bearing compounds seems consistent with the nitrogen-rich nature of the ejecta.

All in all, $\eta$ Car gives us an unmatched opportunity to examine the lifecycle of dust and molecules in the outskirts of a hot, evolved star in affordable timescales. Further and deeper observations to complete its chemical inventory, along with new chemical models able to reproduce its time-dependent conditions,  will provide an insightful view of the kinetic networks at work, contributing to fill the gaps in our knowledge about this fascinating object.
Finally, we remark that a proper understanding of silicon chemistry in sources like $\eta$ Car is also crucial to assess the relevance of early-type massive stars as factories of silicate dust, a role so far reserved to cooler objects and supernovae.

\begin{acknowledgments}
We thank Dr. Juan Garc\'ia de la Concepci\'on for the useful discussion about the silicon chemistry. We also thank the anonymous referee for insightful comments that improved the quality and clarity of the paper. The research leading to these results has received funding from the European Commission Horizon 2020 research and innovation programme under the grant agreement no. 863448 (NEANIAS). J.R.R. acknowledges support by grant PID2019-105552RB-C41 funded by MCIN/AEI/10.13039/501100011033. This paper makes use of the following ALMA data: ADS/JAO.ALMA\#2019.1.00195.L. ALMA is a partnership of ESO (representing its member states), NSF (USA) and NINS (Japan), together with NRC (Canada), MOST and ASIAA (Taiwan), and KASI (Republic of Korea), in cooperation with the Republic of Chile. The Joint ALMA Observatory is operated by ESO, AUI/NRAO and NAOJ.
\end{acknowledgments}

\vspace{5mm}

\subparagraph{Data availability}. The data presented in this work is publicly accessible from the ALMA archive. The NEANIAS project fully supports Open Science and encourages the application of FAIR principles. Therefore,  we have exported the \texttt{MADCUBA} outputs in VO-compliant format \citep{rizzo_j_r_2022_7046117} to ensure interoperability and reproducibility of the modelling results, making them available along with the processed spectra in the Zenodo repository at \url{https://pending..upon..acceptance}. The physical parameters are written as VOTables, and the resulting synthetic spectra as verified FITS files\footnote{\url{https://cdsarc.cds.unistra.fr/vizier.submit/fitsvalidator.html}} compliant with the ObsCore data model\footnote{\url{https://www.ivoa.net/documents/ObsCore/}}.

\bibliography{references}{}
\bibliographystyle{aasjournal}

\appendix

\section{Full spectra} \label{sec:full-spectra}

\begin{figure*}[h]
    \centering
    \figurenum{5}
    \includegraphics[width=0.8\textwidth]{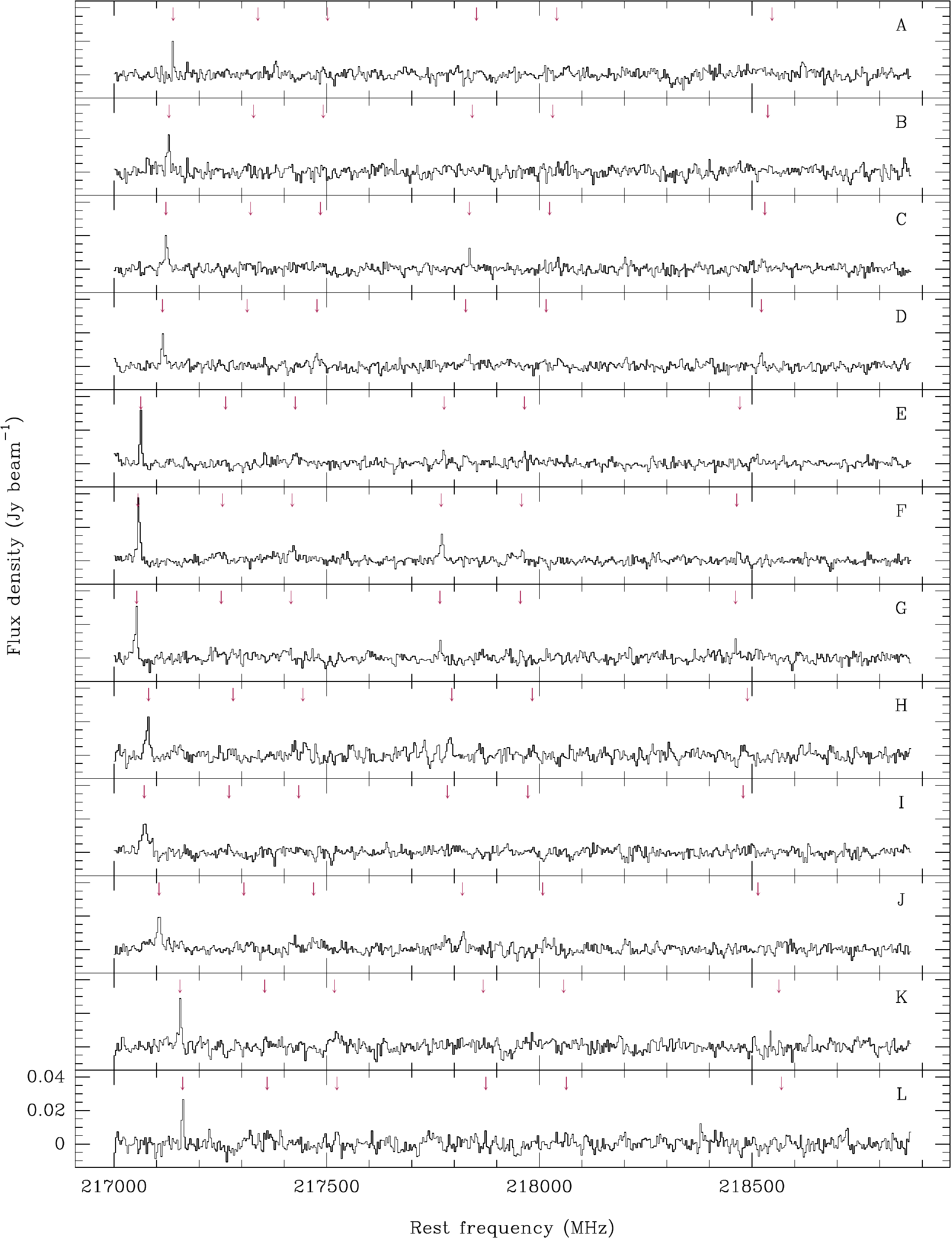}
    \caption{Beam-averaged spectra of $\eta$ Car between 217.000 and 218.850 GHz extracted from the clumps indicated in Figure \ref{fig:maps}. Spectra have been smoothed to a resolution of $\sim$ 5 km s$^{-1}$. In each panel, the markers indicate the frequencies of the transitions listed in Table \ref{tab:mol_inventory}, shifted to the velocity of the strongest SiO component.}
    \label{fig:fig-spw25-appendix}
\end{figure*}

\pagebreak

\begin{figure*}[t]
    \centering
    \figurenum{6}
    \includegraphics[width=0.8\textwidth]{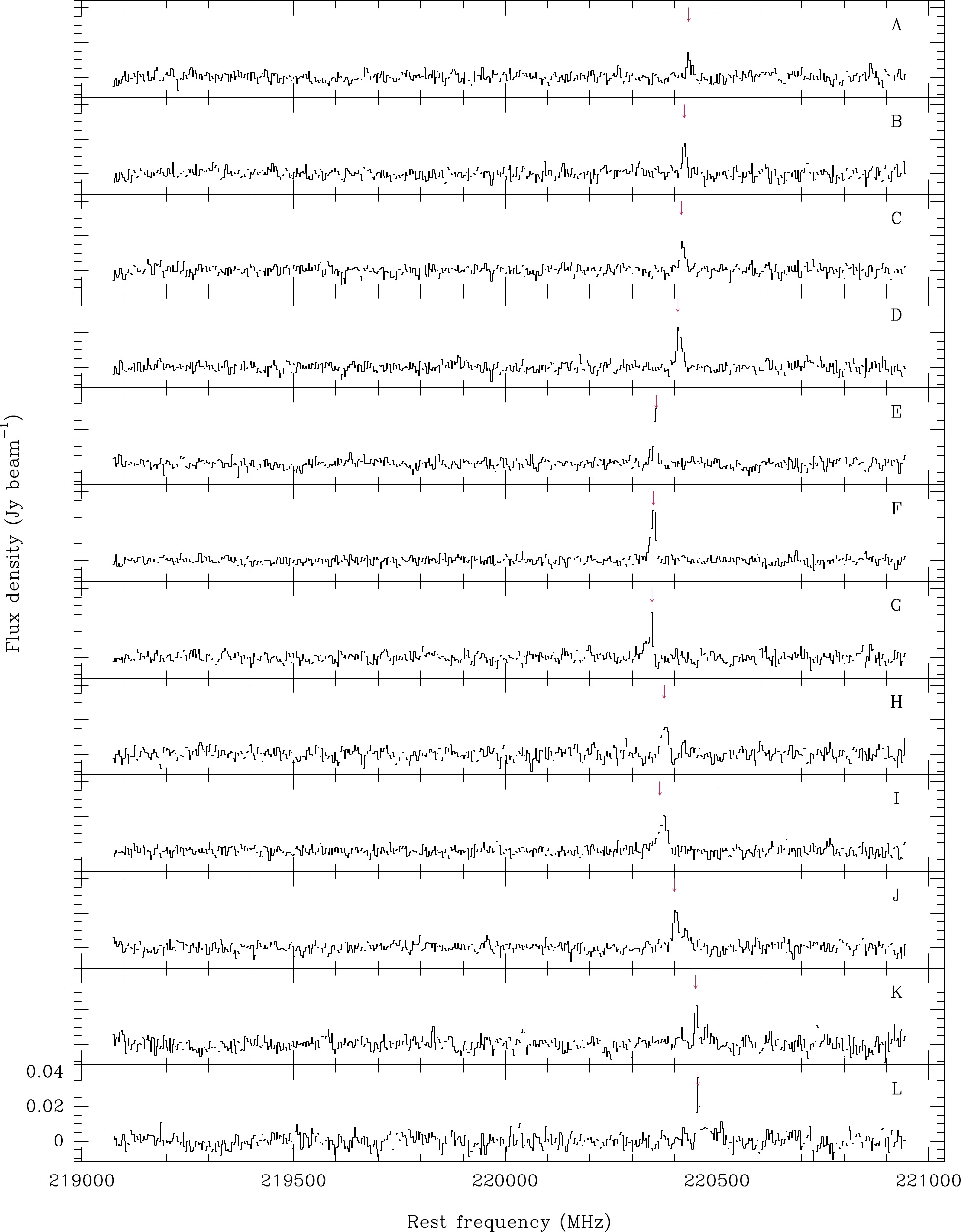}
    \caption{Beam-averaged spectra of $\eta$ Car between 219.100 and 220.950 GHz, extracted from the clumps indicated in Figure \ref{fig:maps}. Spectra have been smoothed to a resolution of $\sim$ 5 km s$^{-1}$. In each panel, the markers indicate the frequency of the $^{13}$CO $J=2\rightarrow1$ transition, shifted to the velocity of the strongest SiO component.}
    \label{fig:fig-spw27-appendix}
\end{figure*}

\pagebreak

\section{Line fitting} \label{sec:line-fitting}

\begin{deluxetable*}{lllcrccrccrcc}[b!]
\tablecaption{\texttt{MADCUBA} line fitting parameters for SiO, SiS and SiN \label{tab:fitting-SiOSiSSiN}. Symbol $\#$  in the fourth column denotes component number, in order of increasing velocity.}
\tablecolumns{12}
\tablenum{3}
\tablewidth{0pt}
\tabletypesize{\scriptsize}
\tablehead{
\colhead{} & \colhead{} & \colhead{} & \colhead{} &
\multicolumn{3}{c}{SiO} &
\multicolumn{3}{c}{SiS} &
\multicolumn{3}{c}{SiN}
\\
\colhead{Pos} &
\colhead{($\Delta\alpha$, $\Delta\delta$)} &
\colhead{$T_\mathrm{ex}$} &
\colhead{$\#$}  &
\colhead{$v_\mathrm{LSR}$} &  \colhead{$\Delta v$} & \colhead{$\int{T_\mathrm{mb} \delta v}$}  & 
\colhead{$v_\mathrm{LSR}$} &  \colhead{$\Delta v$} & \colhead{$\int{T_\mathrm{mb} \delta v}$}   &
\colhead{\footnotesize{$v_\mathrm{LSR}$}} &  {$\Delta v$} & \colhead{$\int{T_\mathrm{mb} \delta v}$} 
 \\
\colhead{} & 
\colhead{} &
\colhead{(K)} & \colhead{}
& \colhead{(km s$^{-1}$)}  & \colhead{(km s$^{-1}$)} &  \colhead{(K km s$^{-1}$)} &
\colhead{(km s$^{-1}$)}  & \colhead{(km s$^{-1}$)} &  \colhead{(K km s$^{-1}$)} &
\colhead{(km s$^{-1}$)}  & \colhead{(km s$^{-1}$)} &  \colhead{(K km s$^{-1}$)} 
}
\startdata
A        & ($+0.9$, $+2.6$)   & 154 & 1 &  $-$47.4      &     3.9 (0.4)     &     25.11 (2.43)   &  $...$      &   $...$         &    $...$        & $...$   &    $...$        &    $...$   \\
\\
B        & ($+1.2$, $+2.2$)   & 166 & 1 & $-$33.5       &    5.4 (0.7)      &     34.16 (2.62)   &  $...$      &   $...$         &    $...$        & $...$   &    $...$        &    $...$    \\
         &                    &     & 2 & $-$25.4       &    7.1 (1.7)      &     15.09 (2.99)   &  $...$      &   $...$         &    $...$        & $...$   &    $...$        &    $...$    \\
\\
C        & ($+0.9$, $+1.9$)   & 180 & 1 & $-$24.3       &    6.7 (0.6)      &     42.52 (3.33)   & $-$24.9      &   4.4 (0.3)   &     17.36 (0.88)   & $...$   &    $...$        &    $...$   \\
\\
D        & ($+1.2$, $+1.7$)   & 185 & 1 & $-$23.5       &    2.5 (0.4)      &     6.17 (0.91)   & $-$24.4   &   5.5 (0.7)   &     10.72 (1.13)   &    $...$     & $...$          & $...$     \\
         &                    &     & 2 & $-$12.4       &    6.4 (0.3)      &     34.66 (1.47)   & $-$12.4   &   2.7 (0.4)   &     5.89 (0.79)   & $-$14.3 &    4.2 (3.8)  &     3.88 (1.45) \\
\\
E        & ($+2.0$, $+0.4$)   & 179 & 1 & 57.8          & 5.9 (0.2)         &      53.86 (1.40)   & 58.0      &   8.6 (1.2)   &     26.13 (3.19)   &    $...$     & $...$          & $...$     \\
\\
F         & ($+2.0$, $-0.2$) & 173 & 1 & 59.2          & 4.8 (0.9)         &     19.35 (2.04)   & 60.0      &   6.0 (1.9)   &      14.49 (2.30)  & 56.6  &    10.0 (4.3)     &      5.84 (2.70) \\ 
          &                  &     & 2 & 65.9          & 5.7 (0.4)         &     65.46 (2.21)   & 65.8      &   5.0 (1.0)   &     23.58 (2.11)       & 64.7  &     4.9 (2.0)     &    4.19 (1.88) \\
\\
G        & ($+1.8$, $-0.9$)   & 163 & 1 & 71.2          & 4.7 (0.5)         &     37.83 (1.98)   &  69.3        &   7.7 (0.6)       &      27.63 (1.99) & 70.1      &    2.9 (0.8)      &     4.51 (1.32) \\
         &                    &     & 2 & 78.4          & 13.7 (2.5)        &     43.98 (3.37)   &  $...$       &   $...$           &    $...$          & $...$     &    $...$          &    $...$    \\
\\
H        & ($+0.5$, $-1.1$)   & 211 & 1 & 23.6          &3.8 (0.7)          &     9.94 (1.44)    &  $...$       &   $...$       &   $...$           & $...$   &    $...$        &    $...$    \\
         &                    &     & 2 & 34.3          &7.3 (0.6)          &     52.68 (2.00)   &  33.7        &   5.9 (1.0)   &   22.04 (1.65)    & $...$   &    $...$        &    $...$    \\
         &                    &     & 3 & 40.4          &2.9 (1.2)          &     7.08 (1.25)   &  39.5         &   4.0 (1.0)   &   12.94 (1.36)    & $...$   &    $...$        &    $...$ \tablenotemark{a}   \\
         &                    &     & 4 & 45.9          &6.3 (1.0)          &     23.08 (1.86)   &  44.9        &   5.0 (1.0)   &   14.05 (1.51)    & $...$   &    $...$        &    $...$    \\
\\
I        & ($-0.3$, $-2.2$)   & 165 & 1 & 44.0          & 31.0 (2.5)        &       82.44 (5.44)   &  $...$       &   $...$       &   $...$           & $...$   &    $...$        &    $...$    \\
         &                    &     & 2 & 45.5          & 3.2 (1.8)         &       11.71 (1.75)   &  $...$       &   $...$       &   $...$           & $...$   &    $...$        &    $...$    \\
\\
J        & ($-1.3$, $-2.3$)   & 162 & 1 & $-$5.7        & 5.8 (0.9)         &      16.68 (2.47)   &  $-$5.2   &   11.3 (1.4)  &     29.25 (2.76)   & $...$   &    $...$        &    $...$    \\
         &                    &     & 2 &  $-$0.6        & 5.0 (1.0)        &      27.05 (2.29)   &  $...$       &   $...$       &   $...$           & $...$   &    $...$        &    $...$    \\       
\\ 
K        &($-2.3$, $-1.1$)    & 168 & 1 & $-$76.1       & 2.5 (0.7)         &     4.96 (1.05)   &  $...$       &   $...$       & $...$           & $...$   &    $...$        &    $...$    \\  
         &                    &     & 2 & $-$70.0       & 5.5 (0.3)         &     45.99 (1.57)  &  $...$       &   $...$       & $...$           & $...$   &    $...$        &    $...$    \\  
         &                    &     & 3 & $-$61.0       & 5.3 (0.8)         &     13.75 (1.54)  &  $...$       &   $...$       & $...$           & $...$   &    $...$        &    $...$    \\  
\\ 
L        & ($-2.0$, $-0.5$)   & 180 & 1 & $-$78.1       & 5.7 (0.3)         &     49.68 (2.20)   & $-$76.5     &    3.5 (0.6)  & 10.03 (1.52) & $...$   &    $...$        &    $...$    \\  
\enddata
\tablecomments{The fitting of SiN only considers the component at 218.512 GHz.}
\tablenotetext{a}{Tentative detection at 2.5$\sigma$ (see text).}
\end{deluxetable*}

\begin{deluxetable*}{lllcrccrcc}[b!]
\tablecaption{\texttt{MADCUBA} line fitting parameters for $^{13}$CO and $^{13}$CN \label{tab:fitting-COCN}. Symbol $\#$  in the fourth column denotes the component number, in order of increasing velocity.}
\tablecolumns{12}
\tablenum{4}
\tablewidth{0pt}
\tabletypesize{\scriptsize}
\tablehead{
\colhead{} & \colhead{} & \colhead{} & \colhead{} &
\multicolumn{3}{c}{$^{13}$CO} &
\multicolumn{3}{c}{$^{13}$CN}
\\
\colhead{Pos} &
\colhead{($\Delta\alpha$, $\Delta\delta$)} &
\colhead{$T_\mathrm{ex}$} & \colhead{$\#$} &
\colhead{$v_\mathrm{LSR}$} &  \colhead{$\Delta$v} & \colhead{$\int{T_\mathrm{mb} \delta v}$}  & 
\colhead{$v_\mathrm{LSR}$} &  \colhead{$\Delta$v} & \colhead{$\int{T_\mathrm{mb} \delta v}$}
 \\
\colhead{} & 
\colhead{} &
\colhead{(K)} & \colhead{}
& \colhead{(km s$^{-1}$)}  & \colhead{(km s$^{-1}$)} & \colhead{(K km s$^{-1}$)}  &
\colhead{(km s$^{-1}$)}  & \colhead{(km s$^{-1}$)} & \colhead{(K km s$^{-1}$)}
}
\startdata
A       & ($+0.9$, $+2.6$) & 154 & 1 & $-46.4$ & 9.2 (1.0) & 33.74 (3.21) & $...$ & $...$ & $...$ \\
\\
B       & ($+1.2$, $+2.2$)  & 166   & 1 & 	$-43.0$ & 2.7 (0.7) &  5.98 (1.10)  & $...$ & $...$ & $...$ \\
        &                   &       & 2 &	$-38.4$ & 3.4 (1.3) & 6.74 (1.23)   & $...$ & $...$ & $...$  \\
        &                   &       & 3 & 	$-33.3$ & 8.8 (2.9) & 38.02 (1.98)  & $...$ & $...$ & $...$  \\
        &                   &       & 4 & 	$-23.8$ & 3.7 (3.9) & 3.84 (1.29)   & $...$ & $...$ & $...$  \\
\\
C       & ($+0.9$, $+1.9$)  & 180   & 1 &   $-33.4$ & 4.9 (0.0) \tablenotemark{a} & 11.13 (2.65) & $...$ & $...$ & $...$  \\
        &                   &       & 2 &   $-25.2$ & 10.9 (0.9) & 46.22 (3.24) & $...$ & $...$ & $...$  \\
\\
D       & ($+1.2$, $+1.7$)  & 185   & 1 &   $-32.8$ & 4.1 (1.1) & 6.63 (1.75)       & $...$ & $...$ & $...$  \\
        &                   &       & 2 &	$-16.6$ & 18.5 (0.8) & 101.84 (3.72)    & $-12.8$ & 13.5 (3.1) & 0.01 (1.67) \\     
\\
E       & ($+2.0$, $+0.4$)  & 179   & 1 &	$57.9$ & 6.1 (0.2) & 51.37 (1.74)       & $58.2$ & 13.5 (3.2) & 0.01 (1.46) \\
        &                   &       & 2 & 	$63.9$ & 6.7 (0.8) & 23.88 (2.43)       & $...$ & $...$ & $...$  \\
\\
F       & ($+2.0$, $-0.2$)  & 173   & 1 & $56.4$ & 1.1 (0.9) &  0.77 (0.40)         & $...$ & $...$ & $...$  \\
        &                   &       & 2 & $64.5$ & 11.4 (0.3) &  91.38 (2.03)       & $62.3$ & 10.4 (2.0) & 0.01 (1.41) \\
        &                   &       & 3 & $73.5$ & 2.4 (0.0) \tablenotemark{a} & 7.89 (3.43)         & $...$ & $...$ & $...$  \\
        &                   &       & 4 & $80.4$ & 2.4 (0.0) \tablenotemark{a} & 7.33 (3.39)         & $...$ & $...$ & $...$  \\
\\
G       & ($+1.8$, $-0.9$)  & 163   & 1 & $71.1$ & 5.6 (0.4) & 38.45 (1.60)         & $...$ & $...$ & $...$  \\
        &                   &       & 2 & $79.8$ & 8.9 (4.7) & 16.83 (2.01)         & $...$ & $...$ & $...$  \\
        &                   &       & 3 & $91.9$ & 13.0 (3.1) & 26.09 (2.42)        & $...$ & $...$ & $...$  \\
\\
H       & ($+0.5$, $-1.1$)  & 211   & 1 & $12.2$ & 1.8 (0.2) &  3.84 (0.41)         & $...$ & $...$ & $...$  \\
        &                   &       & 2 & $21.7$ & 9.6 (0.7) & 34.49 (2.31)         & $25.3$ & 12.5 (6.1) & 0.01 (2.16) \\
        &                   &       & 3 & $31.7$ & 7.8 (1.5) & 10.68 (1.92)         & $...$ & $...$ & $...$  \\
        &                   &       & 4 & $35.4$ & 18.6 (1.4) & 52.00 (3.36)        & $...$ & $...$ & $...$  \\
        &                   &       & 5 & $44.7$ & 4.3 (0.9) & 6.18 (1.20)          & $...$ & $...$ & $...$  \\
\\
I       & ($-0.3$, $-2.2$)  & 165   & 1 & $19.3$ & 6.5 (0.8) & 17.61 (1.78)         & $...$ & $...$ & $...$  \\
        &                   &       & 2 & $29.9$ & 10.1 (0.6) & 43.72 (2.29)        & $...$ & $...$ & $...$  \\
        &                   &       & 3 & $40.3$ & 18.3 (1.4) & 44.90 (3.05)        & $...$ & $...$ & $...$  \\
        &                   &       & 4 & $52.4$ & 29.3 (4.1) & 29.74 (3.79)        & $...$ & $...$ & $...$  \\
        &                   &       & 5 & $70.0$ & 47.0 (9.3) & 23.84 (5.04)        & $...$ & $...$ & $...$  \\
\\       
J       & ($-1.3$, $-2.3$)  & 162   & 1 & $-37.9$ & 5.8 (0.0) \tablenotemark{a} &  25.12 (2.50)      & $...$ & $...$ & $...$  \\
        &                   &       & 2 & $-6.7$ & 24.7 (1.2) & 124.21 (5.17)       & $...$ & $...$ & $...$  \\
\\
K       &($-2.3$, $-1.1$)   & 168   & 1 & $-121.0$ & 4.9 (0.8) & 10.06 (1.44)       & $...$ & $...$ & $...$  \\
        &                   &       & 2 & $-103.7$ & 7.7 (0.6) & 26.15 (1.80)       & $...$ & $...$ & $...$  \\
        &                   &       & 3 & $-71.5$ & 10.2 (0.4) &  58.91 (2.07)      & $-76.4$ & 14.1 (2.5) & 0.02 (2.31) \\
        &                   &       & 4 & $-26.5$ & 5.6 (0.7) & 13.96 (1.54)        & $...$ & $...$ & $...$  \\
\\
L       & ($-2.0$, $-0.5$) & 180   & 1 & $-154.1$ & 3.8 (0.5) &  16.61 (1.76)       & $...$ & $...$ & $...$  \\
        &                   &       & 2 & $-78.0$ & 8.8 (0.3) & 78.93 (2.68)        & $-79.3$ & 13.3 (3.2) & 0.02 (2.39) \\
\enddata
\tablenotetext{a}{The parameter had to be fixed to allow the fitting to converge simultaneously for all components.}
\end{deluxetable*}


\end{document}